\documentclass{elsart}

\usepackage{amsmath}
\usepackage{amssymb}
\usepackage{graphicx}% Include figure files
\usepackage{dcolumn}% Align table columns on decimal point
\usepackage{bm}% bold math
\def\be{\begin{eqnarray}}
\def\ee{\end{eqnarray}}
\def\ba{\begin{array}}
\def\ea{\end{array}}

\begin{document}
\begin{frontmatter}
\title{Exchange-correlation enhancement of the Land$\acute{\text{e}}$-$g^*$ factor in
integer quantized Hall plateaus}

\author[l1]{G. Bilge\c{c}},
\author[l2]{H. \"Ust\"unel Toffoli}
\author[l3]{A. Siddiki}
and
\author[l1]{I. Sokmen}
\address[l1] {Dokuz Eyl\"{u}l University, Physics Department, Faculty of
  Arts and Sciences, 35160 Izmir, Turkey }
\address[l2] {Middle East Technical University, Physics Department,
  Ankara, 06531, Turkey }
\address[l3]{Mu\~gla University, Physics Department, Faculty of Arts and
  Sciences, 48170-K\"otekli, Mugla, Turkey}

\begin{abstract}
We study the emergent role of many-body effects on a two
dimensional electron gas (2DEG) within the Thomas-Fermi-Poisson
approximation, including both the exchange and correlation
interactions in the presence of a strong perpendicular magnetic
field. It is shown that, the indirect interactions widen the
odd-integer incompressible strips spatially, whereas the
even-integer filling factors almost remain unaffected.

%Deniz Eksi, Trakya University, Department of Physics, 22030 Edirne, Turkey\\
%Fax: +90 711 6891010, Email: D.eksi@trakya.edu.tr\\

%\correspondingauthor{Rein van den Boomgaard}
%\telephone{+31(20)5257560}
%\http{http://www.science.uva.nl/~rein}
%\email{rein@science.uva.nl}

\end{abstract}
\end{frontmatter}

Since the discovery of quantum Hall effect ~\cite{vKlitzing80:494}
much effort has been devoted to understand the peculiar transport
properties of the low dimensional systems in the presence of
(Landau) quantizing strong magnetic fields. In the single
particle, noninteracting electron picture, the two-fold degenerate
Landau states are split only due to the Zeeman effect. The Coulomb
interaction enriched generalization of the single particle picture
introduces the compressible and incompressible fluids as a
consequence of the energy gaps. Namely if the Fermi energy is
pinned one of the spin-split Landau levels, due to high density of
states, a metal-like compressible state is formed, otherwise a
quasi-insulating incompressible state exists. Since the
semi-conducting materials in which the experiments are performed,
have a reduced $g^*$-factor (\emph{i.e.} $\simeq -.44$ for GaAs)
it was quite surprising to observe odd integer quantized Hall
plateaus, which is a direct indication of spin resolved transport.
Soon after the experimental observations, the spin effects were
attributed to indirect interactions that enhances the effective
$g^*$-factor. These many body effects were left untouched in the
pioneering work of Chklovskii et al~\cite{Chklovskii92:4026},
which ended in a considerably large discrepancy between their
non-self-consistent theoretical predictions and experiments
considering high-resolution images of Hall
samples~\cite{Tessmer98:6671,Yacoby99:111,Ahlswede02:165}
demonstrating that the strip widths are several times larger than
the model.

As evidenced by these measurements, the single-particle picture is
not sufficient to describe the behavior of the system. In the
presence of exchange and correlation effects, which stem from
many-body interactions, the spin gap in a two-dimensional electron
system (2DES) is expected to be enhanced compared to the single
particle Zeeman energy~\cite{Stoof95:16}. A strong evidence of
enhanced spin splitting as obtained in several theoretical
treatments~\cite{Stoof95:16,Manolescu96:wuerz,ivan00} is the
enlargement of incompressible strips, visible as plateaus in the
spatial filling factor profile. This enhancement is expected to be
much more pronounced in odd integer Hall
plateaus~\cite{Manolescu96:wuerz} due to polarization effects.
Inclusion of the Coulomb interaction \emph{beyond} the classical
Hartree approximation, i.e. both the exchange and correlation
interactions, is possible within the direct diagonalization
techniques ~\cite{Malet07:115306}, prohibitively demanding for the
systems under investigation ~\cite{siddikispin08:1124} or quantum
Monte-Carlo techniques~\cite{tanatar89:5005}. Another affordable yet
accurate alternative for studying exchange and correlation effects
is the density functional theory formalism
(DFT)~\cite{tanatar89:5005,dreizler:book,Igor06:155314}. The most
common treatment of exchange and correlation in DFT of
spin-polarized systems is the so-called local spin density
approximation (LSDA)~\cite{kohnsham}. The goal of the present paper
is to illustrate the effect of addition of exchange and correlation
on the spin gap through an LSDA-corrected self-consistent
Thomas-Fermi Poisson approximation
(TFPA)~\cite{Guven03:115327,Siddiki03:125315,Lier94:7757,J.Oh97:13519}.
To be clear with LSDA, we note that the exchange part is exact,
however, we use the Tanatar-Ceperly parametrization to describe the
correlation part, of course other parameterizations are also
possible~\cite{Attaccalite}. The Attaccalite parametrization is
shown to be in good agreement with the previous ones, at least for
the systems under consideration~\cite{Esa:03}.\\

\begin{figure}
{\centering
\includegraphics[width=1.\linewidth]{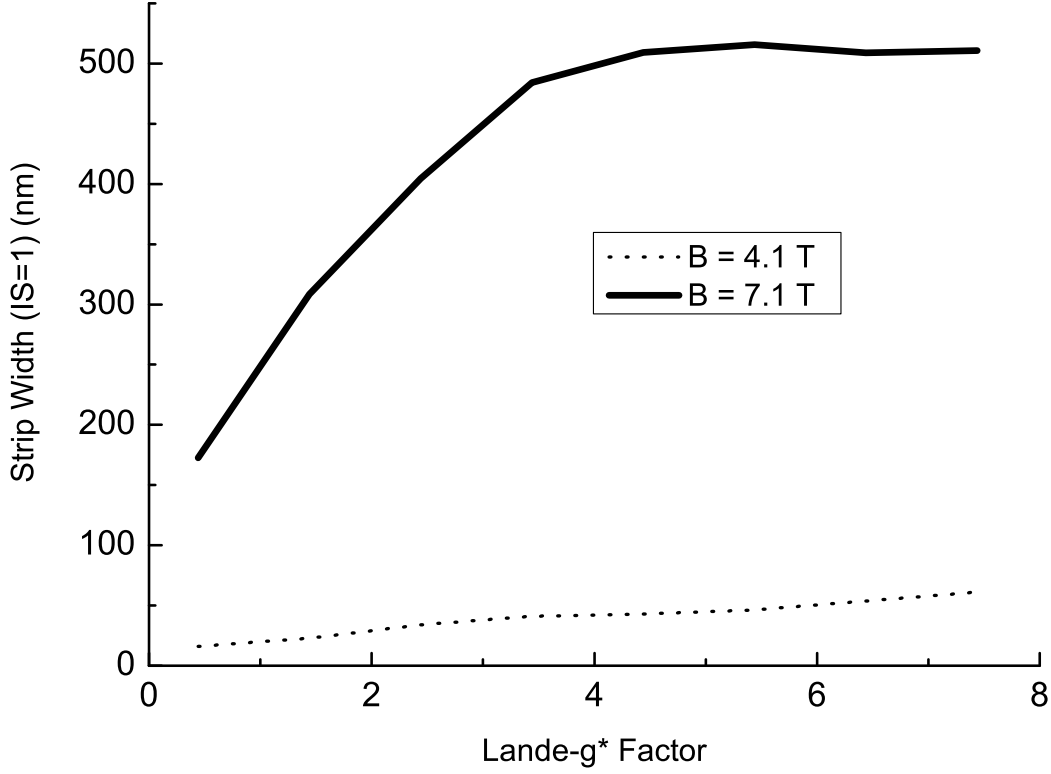}
\caption{The width of the first incompressible strip ($\nu=1$)
without the exchange and correlation as a function of bulk
Lande-$g^*$ factor at $T= 0.05$ K, in a sample of width $3\mu$m, and
for magnetic fields  $B=4.1$ T (dotted line) $B=7.1$ T (solid line).
}\label{fig:fig1}}
\end{figure}
\begin{figure}
\centering
\includegraphics[width=1.\linewidth]{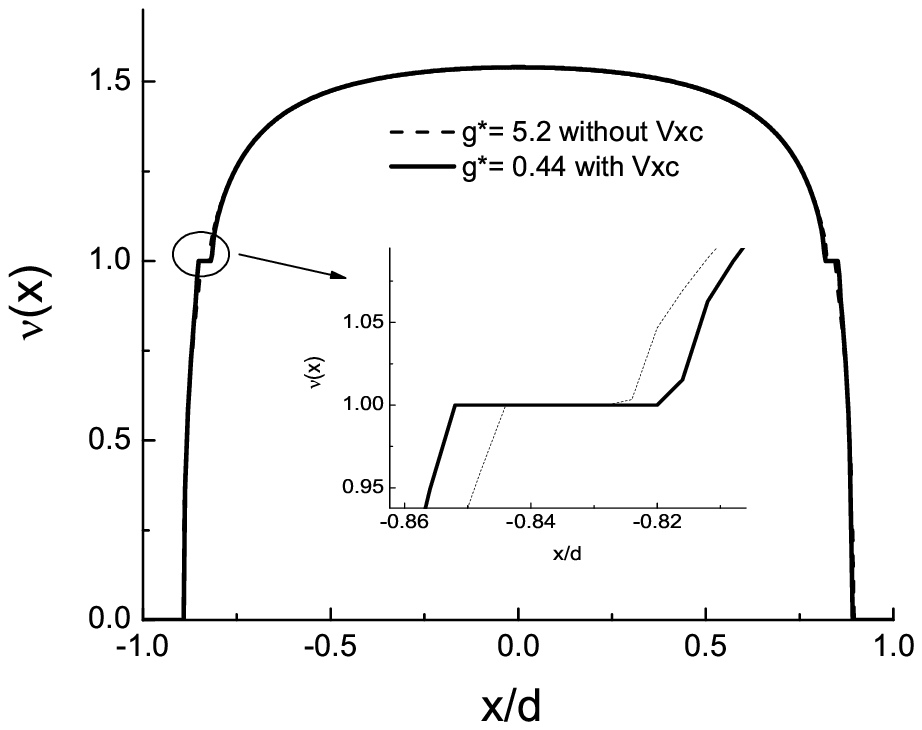}
\caption{The variation of filling factor obtained from
experimental effective Land$\acute{e}-g^*$ Factor
~\cite{siddikispin08:1124} ignoring $V_{\rm xc}$ (dashed line) and
bulk Land$\acute{e}-g^*$ factor including $V_{\rm xc}$ (solid
line). Calculations are performed at default temperature and at
$B=4.7$ T.} \label{fig:fig2}
\end{figure}

We investigate the exchange and correlation interactions in a two
dimensional electron gas confined in a GaAs/AlGaAs
hetero-junction, under the conditions of integer quantized Hall
effect. Spin-split incompressible strips (ISs) with integer
filling factor are first studied using an empirical effective g
factor ~\cite{siddikispin08:1124} then a simplified density
functional approach is utilized to obtain quantitative results. We
consider a two dimensional electron gas (2DEG) with translation
invariance in the $y-$ direction and an electron density $n_{\rm
el}(x)$ confined to the interval $-d<x<d$, in the plane $z=0$.

The Coulomb interaction between electrons is separated into a
classical Hartree and an exchange-correlation potential. The
effective potential is then
\begin{equation}
V(x)=V_{\rm H}(x)+V_{\rm bg}(x)+ V_{\rm Z}+V_{\rm x}(x)+V_{\rm
c}(x). \label{eff-pot}
\end{equation}
The first term in Eq.~\ref{eff-pot} is the Hartree potential,
obtained at each step of the self-consistent TFPA calculations
through the solution of the Poisson equation,
\begin{equation}
V_{\rm
H}(x)=\frac{2e^2}{\bar{\kappa}}\underset{-d}{\overset{+d}\int}dx'n_{\rm
el}(x')K(x,x'), \label{hartree}
\end{equation}
where $-e$ is the electron charge, $\bar{\kappa}=12.4$ is the
average background dielectric constant of GaAs and $K(x,x')$ is the
kernel satisfying the given boundary conditions, $V(-d)=V(d)=0$. In
our study we use the kernel and background potential from
Ref.~\cite{Chklovskii92:4026,Lier94:7757,J.Oh97:13519,Chklovskii93:12605}
\begin{equation}
K(x,x')=\ln\left|\frac{\sqrt{(d^2-x^2)(d^2-x^{'2})}+d^2-x'x}{(x-x')d}\right|.
\label{kernel}
\end{equation}

The background term $V_{\rm bg}(x)$ in Eq.~\ref{eff-pot} describes
the external electrostatic confinement potential composed of gates
and donors modelled by a smooth functional form,
\begin{equation}
V_{\rm bg}(x)=-E_{\rm bg}^0\sqrt{1-(x/d)^2}
\label{background},\quad  E^0_{\rm bg}=2\pi e^2n_0d/\bar{\kappa},
\end{equation}
where $E^0_{\rm bg}$ is the depth of the potential in a positive
background charge density $en_0$. The third term is the Zeeman
energy and reads $V_{\rm Z}=g^*\sigma\mu_BB $, where $g^*$ is the
effective Land$\acute{e}$-g factor, $\mu_B=e\hbar/2m_e$ is the Bohr
magneton and $\sigma=\pm \frac{1}{2}$ is the spin. The last two
terms in Eq.~\ref{eff-pot} are respectively the exchange and
correlation potentials in LSDA. In the present work we use the
Tanatar and Ceperley parametrization ~\cite{tanatar89:5005} with
polarization dependent exchange and correlation potentials. In this
parametrization, $V_{\rm x}(x)$ acts differently on the two spin
channels while $V_{\rm c}(x)$ has a unified form for both channels.

The solution of the TFPA involves the self-consistent
determination of the effective potential given in
Eq.~\ref{eff-pot} for a density
\begin{equation}
n_{\rm el}(x)=\int dE D(E) f(E+V(x)-\mu^*), \label{fermi-dirac}
\end{equation}
obtained in the approximation of a slowly-varying potential valid
in the case in consideration where the magnetic length is larger
than the characteristic length of the potential. Here,
$f(\epsilon)$ is the Fermi function, $D(E)$ and $\mu^*$ are the
density of states (DOS) and the constant equilibrium
electrochemical potential respectively.

In order to motivate the importance of $g^*$-factor enhancement,
we present a preliminary calculation of the first incompressible
strip (IS-1) width that in the presence of only the Zeeman term,
ignoring exchange and correlation. In Fig.~\ref{fig:fig1}, we show
the width of IS-1 while increasing the value of $g^*$ factor as a
free parameter. The width increases significantly until it reaches
a value of approximately 4. For $g^*$ factors larger than this
value the self consistency implies an electrostatic stability
which prevents formation of larger incompressible strips (thick
solid line). However, in lower magnetic fields, the smaller
incompressible strip width of IS-1 width grows approximately
linearly without reaching saturation (thin broken line).
\begin{figure}
\centering
\includegraphics[width=1.\linewidth]{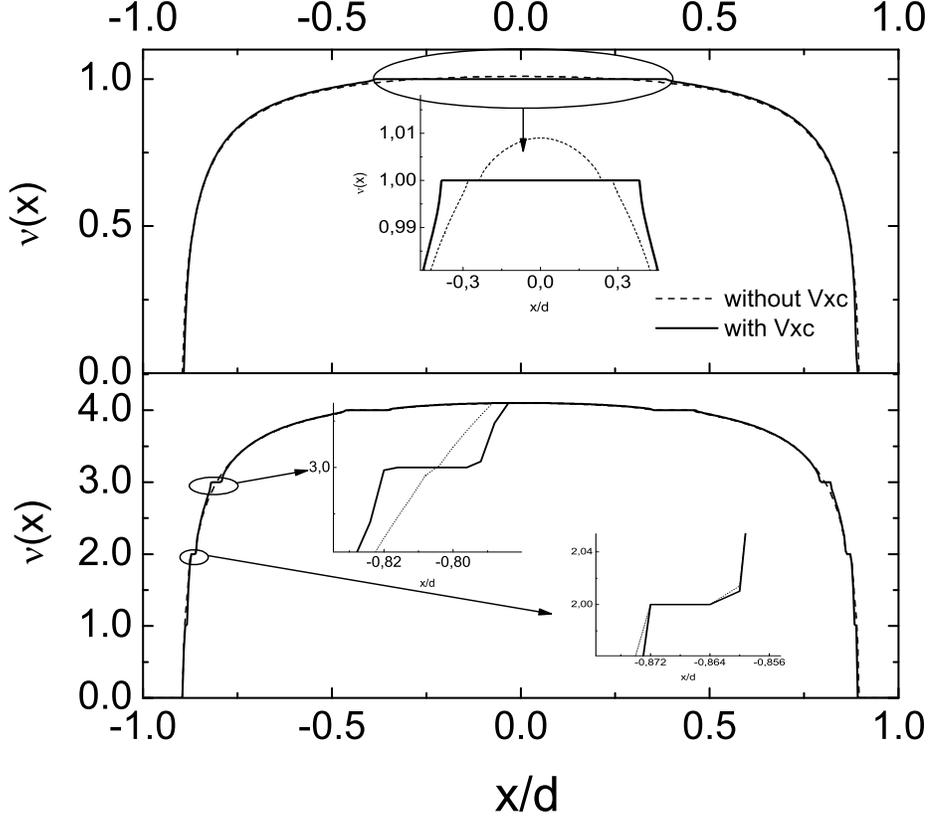}
\caption{Electronic ground state filling factors, neglecting
$V_{\rm xc}$ (dashed line) and including $V_{\rm xc}$ (solid line)
calculated for a sample width of $2d= 3\mu$m, at temperature
$T=0.05$ K and for magnetic fields (a) $B=1.8$ T (b) $B=7.1$ T.}
\label{fig:fig3}
\end{figure}
\begin{figure}
\centering
\includegraphics[width=1.\linewidth]{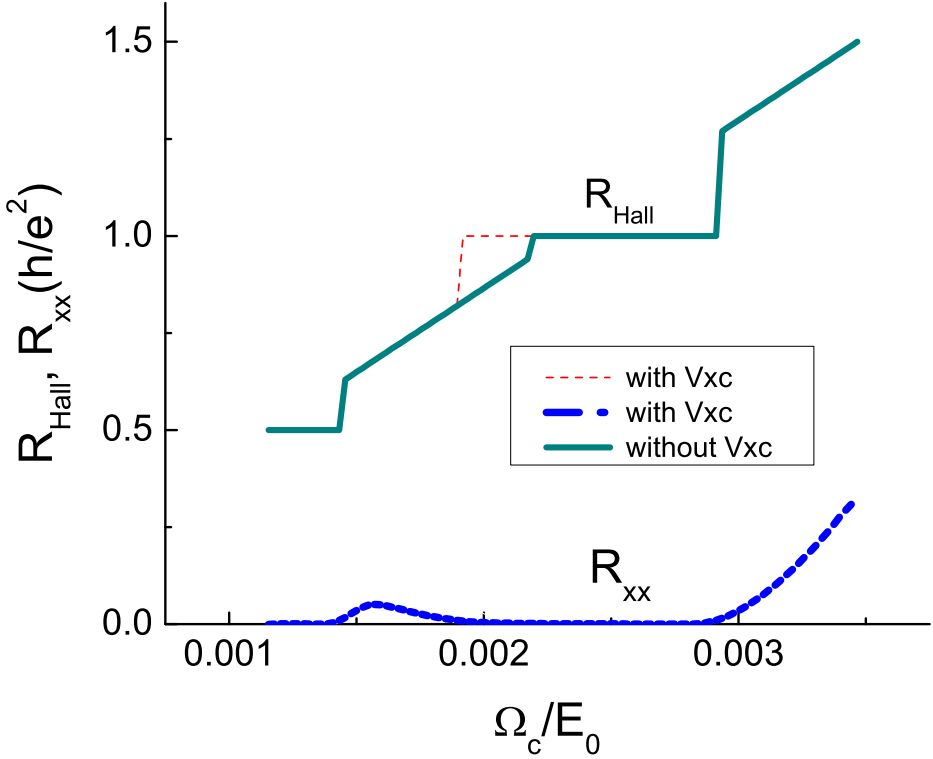}
\caption{Calculated Hall and longitudinal resistances versus scaled
magnetic field $\hbar\omega_c/E_0$, ignoring $V_{\rm xc}$ (solid
line) and including $V_{\rm xc}$ (dashed line). Sample width of $2d=
3\mu$m and for a magnetic field of $B=7.1$ T, at default
temperature.}
 \label{fig:fig4}
\end{figure}
The effect is even more striking when the IS widths are calculated
in the presence and absence of exchange and correlation and
compared, where we fixed the value of $g^*$. In Fig.~\ref{fig:fig2},
we present the local filling factor $\nu(x)$ for the bulk and
experimentally determined $g^*$ factor of
5.2~\cite{siddikispin08:1124}. The figure concludes that, the
inclusion of the indirect interactions spatially enlarges the IS-1
beyond the empirically estimated value of $g^*$, which we attribute
to the incomplete treatment of correlation effects within our
simplified DFT approach.

The filling factor calculated in the presence of the LSDA for a
magnetic field of $B=7.1$ T is displayed in Fig.~\ref{fig:fig3}(a).
At this value of the magnetic field (chosen so as to give a single,
wide incompressible strip) the increase in the strip width in the
presence of $V_{\rm xc}$ is clearly seen. As the magnetic field is
lowered to yield more ISs, the odd-integer strips (IS-1 and IS-3 in
Fig.~\ref{fig:fig3}(b)) continue to be enhanced while those
corresponding to even integers (IS-2 and IS-4 in
Fig.~\ref{fig:fig3}(b)) remain mostly unchanged. This behavior is
due to the nearly full spin polarization for the odd-integer ISs.
Since the exchange-correlation effect often grows with increasing
polarization, its effect is more pronounced for the fully polarized
odd-integer ISs. On the other hand, the even-integer,
spin-compromised ISs are effected only to a small extent.

At a final step we show our transport results obtained within a
local version of the Ohm's law~\cite{siddiki2004} where the local
conductivity tensor entities are assumed to take a simple
analytical form ~\cite{siddikispin08:1124},
$\sigma_l(x)=\frac{e^2}{h}(\nu(x)-[|\nu(x)|])^2$ and
$\sigma_H(x)=\frac{e^2}{h}\nu(x)$. The global resistances are
obtained by utilizing the equation of continuity and translation
invariance in the presence of a fixed imposed external current.
Fig.~\ref{fig:fig4} presents the calculated resistances with and
without including indirect interactions. One can clearly observe
that, the existence of $V_{\rm xc}$ enlarges the $\nu=1$ Hall
plateau drastically, which is exactly the case in the
experiments~\cite{Ahlswede02:165}.

We have calculated the filling factor profile of 2DESs in the
presence of a strong magnetic field using the self-consistent
TFPA. The exchange-correlation potential, included within the
Tanatar-Ceperley parametrization of LSDA is observed to enhance
the IS widths at integer filling. Our method provides a fully
self-consistent calculation scheme to obtain even and odd integer
quantized Hall plateaus, displaying clear differences in width
enhancement due to spin polarization. The results indicate that
the enhancement effect is much more pronounced in odd-integer
fillings due to the possibility of polarization while the
even-integer, spin-compromised plateaus are hardly affected. The
distinguishing part of this work relays on the fact that, without
any complicated numerical (e.g. parallel computing) or analytical
(e.g. localization) methods we can obtain the odd integer
quantized Hall plateaus in a good qualitative agreement with the
experiments.

%\bibliography{zitate}
%\bibliographystyle{elsart-num}
%\bibliography{cite}
%\bibliography{siddiki}

\begin{thebibliography}{10}

\bibitem{vKlitzing80:494}
K. v.~Klitzing, G. Dorda and M. Pepper, Phys. Rev. Lett. {\bf 45},  494
  (1980).

\bibitem{Chklovskii92:4026}
D.~B. Chklovskii, B.~I. Shklovskii and L.~I. Glazman, Phys. Rev. B {\bf 46},
  4026  (1992).

\bibitem{Tessmer98:6671}
S.~H. Tessmer, P.~I. Glicofridis, R.~C. Ashoori, L.~S. Levitov and M.~R.
  Melloch, Nature {\bf 51},  6671  (1989).

\bibitem{Yacoby99:111}
A. Yacoby, H.~F. Hess, T.~A. Fulton, L.~N. Pfeiffer and K.~W. West, Solid State
  Communications {\bf 111},  1  (1999).

\bibitem{Ahlswede02:165}
E. Ahlswede, J. Weis, K. von Klitzing and K. Eberl, Physica E {\bf 12},  165
  (2002).

\bibitem{Stoof95:16}
T.~H. Stoof and G.~E.~W. Bauer, Phys. Rev. B {\bf 52},  16  (1995).

\bibitem{Manolescu96:wuerz}
A. Manolescu and R.~R. Gerhardts,  in {\em Proc. 12th Intern. Conf. Appl. High
  Magnetic Fields, {W\"urzburg}, 1996} (World Scientific, Singapore, 1997),
  this volume.

\bibitem{ivan00}
I.~A. Larkin and L.~S. Levitov, Physica E {\bf 6},  91  (2000).

\bibitem{Malet07:115306}
F. Malet, M. Pi, M. Barranco, L. Serra and E. Lipparini, Phy.Rev.B {\bf 76},
  115306  (2007).

\bibitem{siddikispin08:1124}
A. Siddiki, Physica E {\bf 40},  1124  (2008).

\bibitem{tanatar89:5005}
B. Tanatar and D.~M. Ceperley, Phys.Rev.B {\bf 39},  5005  (1989).

\bibitem{dreizler:book}
R.~M. Dreizler and E.~K.~U. Gross, {\em Density Functional Theory} (Springer,
  Berlin, 1990).

\bibitem{Igor06:155314}
S. {Ihnatsenka} and I.~V. {Zozoulenko}, Phys. Rev. B {\bf 73},  155314  (2006).

\bibitem{kohnsham}
W. Kohn and L. Sham, Phys. Rev. {\bf 140},  A1133  (1965).

\bibitem{Guven03:115327}
K. G{\"u}ven and R.~R. Gerhardts, Phys. Rev. B {\bf 67},  115327  (2003).

\bibitem{Siddiki03:125315}
A. Siddiki and R.~R. Gerhardts, Phys. Rev. B {\bf 68},  125315  (2003).

\bibitem{Lier94:7757}
K. Lier and R.~R. Gerhardts, Phys. Rev. B {\bf 50},  7757  (1994).

\bibitem{J.Oh97:13519}
J.~H. Oh and R.~R. Gerhadts, Phys.Rev. B {\bf 56},  13519  (1997).

\bibitem{Attaccalite}
C. Attaccalite, S. Moroni, P. Gori-Giorgi and G.~B. Bachelet, Phys. Rev. Lett.
  {\bf 88},  256601  (2002).

\bibitem{Esa:03}
H. Saarikoski, E. R\"as\"anen, S. Siljam\"aki, A. Harju, M.~J. Puska and R.~M.
  Nieminen, Phys. Rev. B {\bf 67},  205327  (2003).

\bibitem{Chklovskii93:12605}
D.~B. Chklovskii, K.~A. Matveev and B.~I. Shklovskii, Phys. Rev. B {\bf 47},
  12605  (1993).

\bibitem{siddiki2004}
A. Siddiki and R.~R. Gerhardts, Phys. Rev. B {\bf 70},  195335  (2004).

\end{thebibliography}
%\bibliographystyle{prsty}

\end{document}